\documentclass[english,twoside,a4paper,10pt]{article}

\usepackage[latin1]{inputenc}
\usepackage[T1]{fontenc}
\usepackage{amsmath}
\usepackage{amsfonts}
\usepackage{graphicx}
\usepackage{subfigure}
\usepackage{a4wide}
\usepackage{amssymb}
\usepackage{fancyhdr}
\usepackage{mathrsfs}
\usepackage[toc,page]{appendix}

\begin{document}

\title{\bf Bounce solutions in viscous fluid cosmology}
\author{ 
Ratbay Myrzakulov\footnote{Email address: rmyrzakulov@gmail.com},\,\,\,
Lorenzo Sebastiani\footnote{E-mail address: l.sebastiani@science.unitn.it
}\\
\\
\begin{small}
Eurasian International Center for Theoretical Physics and  Department of General
\end{small}\\
\begin{small} 
Theoretical Physics, Eurasian National University, Astana 010008, Kazakhstan
\end{small}\\
}

\date{}

\maketitle


\begin{abstract}
We investigate the bounce cosmology induced by inhomogeneous viscous fluids in FRW space-time (non necessarly flat), taking into account the early-time acceleration after the bounce. 
Different forms for the scale factor and several examples of fluids will be considered.
We also analyze the relation between bounce and finite-time singularities and between the corresponding fluids realizing this scenarios. In the last part of the work, the study is extended to the framework of $f(R)$-modified gravity, where the modification of gravity may also be considered as an effective (viscous) fluid producing the bounce. 
\end{abstract}



\tableofcontents
\section{Introduction}

The cosmological observations reveal that the universe is expanding in an accelerated way~\cite{WMAP}.
Apart the introduction of small and positive
Cosmological Constant in the framework of General Relativity, where the acceleration is induced by the negative pressure of the dark energy with Equation of State parameter 
$\omega=-1$,
several descriptions of the cosmic acceleration have been presented in the recent literature,
making the dark energy issue the ``Mystery of the Millennium''~\cite{Pad}, with many implications in fundamental physics. 
The data constrain $\omega$ to be very close to minus one, but in principle different
forms of dark fluid (phantom, quintessence...), satisfying a suitable Equation of State are allowed.
Moreover, the existence of an accelerated epoch also in the early-time universe, namely the inflation,
which cannot be driven by standard matter/radiation,
adds some new interest in the investigation of general forms, behaviours and solutions of dark fluids. 
We also observe that, despite to the fact that many macroscopic physical systems, like the large scale structure of matter, can be approximated like perfect fluids (with Equation of State parameter $\omega=\text{const}$), we cannot exclude non-perfect fluid representation for the dark components of the universe (whose origin remains unknown) like inhomogeneous and/or viscous fluid representation.  Namely, the Equation of Steate parameter of the dark energy may be not a constant or its pressure may depend on the expansion rate of the universe due to some viscosity.
The investigation of such a kind of fluids is motivated by several reasons.  
For example, in the last years the interest in modified theories of gravity,
where some combination of curvature invariants (Riemann tensor, Weyl tensor, Ricci tensor and so on) replaces or is added into the classical Hilbert-Einstein action of General Relativity,
has grown up (see Refs.~\cite{Review-Nojiri-Odintsov} --\cite{SebRev} and 
Ref.\cite{others}), and  
it is worth considering that this theories have a corresponding description in the fluid-like form, so that the study of inhomogeneous viscous fluids is one of the easiest way to understand some of the general aspects of modified theories also (for a recent review of inhomogeneous fluids as an equivalent description of different 
theoretical models, see Ref.~\cite{revf}).

When we consider universe contents different from the standard matter ones, we may find several interesting cosmological solutions with a great varieties of features like oscillations, singularities etc. Among them, the bounce solutions (where a cosmological contraction is followed by an expansion at a finite time) 
are interesting to analyze (see Refs.~\cite{Novello} for a review).
The idea that, instead from an initial singularity, the universe has emerged from a cosmological bounce furnishes an alternative scenario to the Big Bang theory. In the so-called matter bounce scenario~\cite{matterbounce}, in the initial contraction the universe is in a matter-dominated stage, after that a bounce without any singularity appears, and the expanding universe 
with the correct observed matter spectrum is generated: in such a case, the precision of the anisotropies predicted by the model are well confirmed by the observations of the Cosmic Microwave Background (CMB) anisotropies.
Many different aspects of bounce cosmology have been analyzed in the literature
(see Ref.~\cite{inst} for BKL instability, Ref.~\cite{ek}
for the Ekpyrotic scenario, 
Ref.~\cite{add}
for
confrontation of bounce universe
with Planck observations
and Refs.~\cite{add2, otbounce} for other works). Finally, in the recent work of Ref.~\cite{Odbounce}, the bounce solutions have been considered in the framework of modified gravity and massive bigravity. 

The aim of this work is to investigate the bounce cosmology induced by inhomogeneous viscous fluids in Friedmann-Robertson-Walker space-time (not necessarly flat). 
We will discuss different bounce solutions and the feautures of the related dark fluids, taking into account the necessity to have a cosmic (inflationary) acceleration after the bounce. In particular, we are interested in the relation between bounce and singular solutions, and in the corresponding relation between the dark fluids realizing such scenarios. In the last part of the work, we also will extend the study to $f(R)$-modified gravity, where the modification to Einstein's gravity can be viewed as an (effective) viscous fluid: in this case, the realization of bounce solutions will be considered in flat FRW space-time.  

The paper is organized as follows. In Section {\bf 2}, the formalism of inhomogeneous viscous fluids in FRW universe is presented. In Section {\bf 3}, we will analyze the bounce solutions in fluid cosmology with exponential scale factor, we will study the feauture of the models and we will analyze the appearance of singular solutions. Some explicit examples of fluid realizing such a bounce will be presented. In section {\bf 4}, the same investigation will be carried out for the bounce solutions with power-law scale factor. Section {\bf 5} is devoted to bounce solutions in $f(R)$-gravity by using a fluid-like representation.
Conclusions and remarks are given in Section {\bf 6}.

We use units of $k_{\mathrm{B}} = c = \hbar = 1$ and denote the
gravitational constant, $G_N$, by $\kappa^2\equiv 8 \pi G_{N}$, such that
$G_{N}^{-1/2} =M_{\mathrm{Pl}}$, $M_{\mathrm{Pl}} =1.2 \times 10^{19}$ GeV being the Planck mass.


\section{Inhomogeneous viscous fluids in FRW space-time.}

Let us start by recalling the Friedmann Equations of Motion (EOMs) for the Friedmann-Robertson-Walker (FRW) metric in spherical coordinates $r,\theta,\phi$,
\begin{equation}
ds^{2}=-dt^{2}+a^{2}(t)\left[\frac{dr^2}{1-k^2r^2}+r^2 d\Omega^2\right]\,,\quad d\Omega^2=(d\theta^2+\sin^2\theta d\phi^2)\,,\label{metric}
\end{equation}
which read
\begin{equation}
H^2+\frac{k}{a^2}=\frac{\kappa^2\rho}{3}\,,\quad -\frac{(2\dot H+3H^2)}{\kappa^2}=p\,.\label{EOMs}
\end{equation}
In the above expressions, $a(t)$ is the scale factor of the universe ($r=r'/\sqrt{|a(t)|^2}$, $r'$ being the physical radial coordinate), $k=-1,0,1$ is the spatial curvature which correspons to the hyperbolic, flat or spherical space, respectively, and 
$H=\dot a(t)/a(t)$ is the Hubble paramter where the dot denotes the derivative with respect to the cosmological time $t$.  
The cosmological parameter reads
\begin{equation}
\Omega=1+\frac{k}{a^2 H^2}\,,\label{cosmpar}
\end{equation}
and in general can be different to one. 

In the Friedmann equations, $p$ and $\rho$ are the pressure and the energy density of the fluid contents of the universe which must satisfy the conservation law,
\begin{equation}
\dot\rho+3H(\rho+p)=0\,.\label{CL}
\end{equation} 
In this work, we will consider the general form for the Equation of State (EoS) of inhomogeneous viscous fluids, namely ~\cite{fluidsOd2, Od3, Od4}
\begin{equation}
p=\omega(\rho)\rho-B(a(t),H, \dot{H}...)\,,\label{start}
\end{equation}
where the EoS parameter, $\omega(\rho)$, may depend on the energy density, and the bulk viscosity  $B(a(t),H, \dot{H}...)$ is a general function of the scale factor, the Hubble parameter and its derivatives. 
On thermodynamical grounds, in order to have the positive sign of the entropy change in an irreversible process, the bulk viscosity must be a positive quantity~\cite{Alessia, Alessia(2)}. 
The stress-energy tensor of fluid $T_{\mu\nu}$ is given by
\begin{equation}
T_{\mu\nu}=\rho u_{\mu}u_{\nu}+\left[\omega(\rho)\rho+B(\rho,a(t),H, \dot{H}...)\right](g_{\mu\nu}+u_{\mu}u_{\nu})\,, 
\end{equation}
where $u_{\mu}=(1,0,0,0)$ is the four velocity vector. 
The fluid energy conservation law (\ref{CL}) finally leads to
\begin{equation}
\dot{\rho}+3H\rho(1+\omega(\rho))=
3 H B(\rho,a(t),H, \dot{H}...)
 \label{conservationlawfluid}\,.
\end{equation}
In the next sections, we will revisit some simple bounce solutions discussing the general features of the fluids which realize them and the possibility to have an early-time acceleration after the bounce. Some explicit examples of this fluids will be furnished.

\section{Bounce solutions with exponential scale factor}

We start by examining the bounce scenario where the scale factor and therefore the Hubble parameter behave as 
\begin{equation}
a(t)=a_0\text{e}^{\alpha (t-t_0)^{2n}}\,,\quad H(t)=2n\alpha\,(t-t_0)^{2n-1}\,,\quad n=1,2,3...
\label{exp}
\end{equation}
where $a_0\,,\alpha$ are positive (dimensional) constants and $n$ is a positive natural number from which depends the feature of the bouncing. Moreover, $t_0>0$ is the fixed bounce time. When $t<t_0$, the scale factor decreases and we have  a contraction with negative Hubble parameter, at $t=t_0$ we have the bounce, such that $a(t=t_0)=a_0$, and when $t>t_0$ the scale factor increases and the universe expands with positive Hubble parameter. 

It is worth to spending some words on the case of $n$ non positive natural number. First of all, the special choice $n=1/2$ corresponds to the de Sitter solution ($H(t)=\text{const}$). 
Then, when $n<1/2$ or $n$ is a positive non-natural number, the bounce is changed in a finite-time singularity occurring at $t=t_0$. It means, that the Hubble parameter or some of its derivatives (and therefore the curvature) diverge at that time, and we may describe two different expansion (or contraction) cosmological histories, but the two branches with $t>t_0$ ($H(t)<0$) and $t<t_0$ ($H(t)>0$) are not connected respect to each other: in other words, the universe starts or finish with a singularity. In the specific, the finite-time singularities can be classified in the following way~\cite{classificationSingularities}:
\begin{itemize}
\item Type I~\cite{Caldwell}--\cite{Rip7}: for $t\rightarrow t_{0}$, $a(t), H(t), \dot H(t)\rightarrow\infty$, namely the scale factor, the effective energy density and the effective pressure of the universe diverge.
It corresponds to the case $n<0$.
\item Type II (sudden \cite{sudden, suddenOd}):
for $t\rightarrow t_{0}$, $a(t)\rightarrow\text{const}$,
$H(t)\rightarrow\text{const}$ and $\dot H(t)
\rightarrow\infty$, namely the effective pressure of the universe diverges.
It corresponds to the case $0<n<1/2$.
\item Type III~\cite{IIIOd}: for $t\rightarrow t_{0}$, $a(t)\rightarrow \text{const}$,
$H(t)\rightarrow\infty$ and
$|\dot H(t)|\rightarrow\infty$, namely the effective energy density and pressure of the universe diverge.
It corresponds to the case $1/2<n<1$.
\item Type IV~\cite{classificationSingularities}: for $t\rightarrow t_{0}$, only the higher derivatives of $H(t)$ diverge.
It corresponds to the case
$n>1$, but $n\neq m/2$, where $m$ is an integer number.
\end{itemize}
Finally, when $n=m/2$, $m$ odd integer number, the scale factor possesses a saddle point at $t=t_0$, but the bounce is absent.

The presence of an initial singularity suggests the Big Bang scenario, while the bounce solution brings to a contracting/expanding universe. In such a case, 
\begin{equation}
\frac{\ddot a}{a}=H^2+\dot H=
2n\alpha(t-t_0)^{2(n-1)}\left[2n\alpha(t-t_0)^{2n}+(2n-1)
\right]\,,
\end{equation}
and we have an acceleration.  In particular, after the bounce, the universe expands in an accelerated way and the inflationary scenario may be suggested.

Let us return to the bounce solution (\ref{exp}). From the first EOM in (\ref{EOMs}) we obtain
\begin{equation}
\rho=\frac{3}{\kappa^2}\left[4n^2\alpha^2(t-t_0)^{2(2n-1)}+\frac{k}{a_0^2\text{e}^{2\alpha(t-t_0)^{2n}}}\right]\,.\label{dd}
\end{equation}
It is easy to see that for the flat universe ($k=0$) or for the spherical universe ($k=1$), this quantity is positive defined, but in the hyperbolic space ($k=-1$) we have a region where the fluid possesses a negative energy density (in particular, $\rho=-3/(a_0\kappa)^2$ at $t=t_0$).
For this reason, we will concentrate on the first two (physical) cases. 

In the flat universe,  the energy density of fluid decreases with the contraction,  is equal to zero at $t=t_0$, and increases with the subsequent expansion.

In the spherical universe, if $n>1$, there is a region around the bounce where the energy density of fluid increases during the contraction and decreases during the expansion, reaching the value of $\rho=3/(a_0\kappa^2)$ at $t=t_0$. 
This is clear if we analyze the time derivative of the energy density,
\begin{equation}
\dot\rho=\frac{3}{\kappa^2}\left[8n^2(2n-1)\alpha^2(t-t_0)^{4n-3}-4n\alpha(t-t_0)^{2n-1}\frac{k}{a_0^2\text{e}^{2\alpha(t-t_0)^{2n}}}\right]\,,\label{dot}
\end{equation}
for $k=1$. Let us consider $n>1$. We see that, when $t\ll t_0$, since $(t-t_0)<0$ and $(t-t_0)^{4n-3}\gg (t-t_0)^{2n-1}$, the first negative term is dominant and the energy density decreases. However, when $t$ approaches to $t_0$, the second positive term becomes dominant 
and the energy density starts to increase until $t_0$ where has a local maximum. After that, energy density decrases as soon as $t$ remains close to $t_0$, but, when $t\gg t_0$, the energy density increases again. This mechanism is interesting if we consider that after the bounce the cosmological parameter (\ref{cosmpar}), namely
\begin{equation}
\Omega=1+\frac{k}{a_0^2\alpha^2(t-t_0)^{2(2n-1)}\text{e}^{2\alpha(t-t_0)^{2n}}}\,,
\end{equation}
decreases as a consequance of the acceleration of the universe. In this way, the solution (\ref{exp}) with $n>1$ may give rise to an accelerated universe, whose energy density decreases with the cosmological parameter. This scenario may be compatible with the inflation, whose effective energy density decreases making possible the exit from this period, at the end of whose the cosmological parameter is close to one: it is clear that to reproduce such comology other fluid contents which become dominant during this phase must be added to the model to produce a subsequent deceleration.  

In the case $n=1$ with spherical geometry, expression (\ref{dot}) is equal to zero only for $t=t_0$: the energy density of fluid decreases before the bounce and always increases after that.

We will analyze now the fluids producing the bounce using the conservation law (\ref{conservationlawfluid}).
In order to do it, we must consider the general form (\ref{start}) of such a kind of fluids, since it is well known that standard perfect fluids with constant EoS parameter cannot reproduce this kind of cosmology. Firstly, we will investigate inhomogeneous non-viscous fluids, and afterwards we will introduce the viscosity.

\subsection{Fluids realizing the bounce with exponential scale factor}

As a first example, we analyze the case of non viscous fluids, namely $B(a(t),H, \dot{H}...)=0$ in (\ref{start}). In the simplest case of $k=0$ (flat topology), we immediatly have from the conservation law, 
\begin{equation}
p=-\rho-\rho^{\frac{(n-1)}{(2n-1)}}\left[
\frac{3}{\kappa^2}(2n\alpha)^2
\right]^{\frac{2n}{4n-2}}\left(\frac{2n-1}{3n\alpha}\right)\,,
\end{equation}
and the EoS paramter reads
\begin{equation}
\omega(\rho)=-1-\rho^{\frac{-n}{(2n-1)}}\left[
\frac{3}{\kappa^2}(2n\alpha)^2
\right]^{\frac{2n}{4n-2}}\left(\frac{2n-1}{3n\alpha}\right)\,.\label{oo}
\end{equation}
This kind of fluid in flat FRW space-time has been often analyzed in the literature. In particular,
in Refs.~\cite{NO1, NO2, mioultimo} its behaviour connected with the presence of singularities has been discussed. As we stressed at the beginning of the Chapter, the occurrence of the bounce scenario for positive integer values of $n$ finds some correspondence in the emerging of singular solutions in the same fluid models where $n$ is negative or positive non integer number. In the specific, if the power law of $\rho$ in (\ref{oo}) is negative, the bounce is realized, but if it is positive, a singularity appears.

We can also rewrite the fluid by using other forms of Equation of State and by introducing the bulk viscosity. A simple example is given by a constant EoS paramter (here, $\omega=-1$) and a bulk viscosity depending on Hubble paramter only, namely
\begin{equation}
B(a(t),H, \dot{H}...)=3 H\zeta(H)\,,\label{eq.state}
\end{equation}
where $\zeta(H)>0$ is the bulk viscosity. In our specific case, for $k=0$, the fluid Equation of State assumes the form
\begin{equation}
p=-\rho-3H\zeta(H)\,,\quad
\zeta(H)=\left(\frac{3}{\kappa^2}\right)^{\frac{2n-1}{2n-1}}
\left(
2n\alpha
\right)^{\frac{1}{2n-1}}\left(\frac{2n-1}{3}\right)H^{-\frac{1}{2n-1}}\,.
\label{p1}
\end{equation}
When we introduce the spatial curvature and $k\neq 0$, the EoS of the fluid becomes more complicate and we need a viscosity depending on the scale factor also. A simple formulation for solution (\ref{exp}) is 
\begin{equation}
p=-\rho-3H\zeta(H, a(t))
\,,
\end{equation}
where
\begin{equation}
\zeta(H,a(t))=\left(\frac{3}{\kappa^2}\right)^{\frac{2n-1}{2n-1}}
\left(
2n\alpha
\right)^{\frac{1}{2n-1}}\left(\frac{2n-1}{3}\right)H^{-\frac{1}{2n-1}}
-\frac{2k}{(3H)\kappa^2a(t)^2}\,.
\end{equation}
When the scale factor becomes large, this expression coincides with (\ref{p1}) and we can treat the viscous fluid like a fluid in the flat space.

\section{Bounce solutions with power-law scale factor}

In this Section, we will analyze the following form for the scale factor and Hubble paramter,
\begin{equation}
a(t)=a_0+\alpha(t-t_0)^{2n}\,,\quad H (t)=\frac{2n\alpha(t-t_0)^{2n-1}}{a_0+\alpha(t-t_0)^{2n}}\,,\quad n=1,2,3...
\label{pow}
\end{equation}
where $a_0$, $\alpha$ are positive (dimensional) constants and $n$ is a positive natural number. The time of the bounce is fixed at $t=t_0$. 
When $t<t_0$, the scale factor decreases and we have  a contraction with negative Hubble parameter, at $t=t_0$ we have the bounce, such that $a(t=t_0)=a_0$, and when $t>t_0$ the scale factor increases and the universe expands with positive Hubble parameter. 

If $a_0=0$ we obtain a bounce solution with singularity, namely the universe contracts until $a(t=t_0)=0$, where the Hubble parameter and therefore the curvature diverge. However, the scale factor does not become singular and starts to increase after $t_0$ realizing the bounce. 

When $n$ is a negative number, the scale factor diverges at $t=t_0$ and we can set $a_0=0$ without loss of generality. In this case, we encounter the so called Big Rip singularity, where $H(t)=-2n/(t_0-t)$, $H$ being positive for $t<t_0$ and diverging with the scale factor at $t=t_0$. This is an important solution of the Friedmann equations in the case of phantom perfect fluids with $\omega<-1$~\cite{Caldwell}, and in fact it is a possible scenario for the dark energy epoch of the universe today.

Let us return to the bounce solution (\ref{pow}). We get
\begin{equation}
\frac{\ddot a}{a}=\frac{2n(2n-1)\alpha(t-t_0)^{2(n-1)}}{a_0+\alpha(t-t_0)^{2n}}\,,
\end{equation}
and we have an acceleration before and after the bounce. 
Moreover, from the first EOM in (\ref{EOMs}) we derive
\begin{equation}
\rho=\frac{3}{\kappa^2\left[a_0+\alpha(t-t_0)^{2n}\right]}\left[\frac{4n^2\alpha^2(t-t_0)^{4n-2}+k}{a_0+\alpha(t-t_0)^{2n}}\right]\,.\label{dd}
\end{equation}
It is easy to see that in the flat ($k=0$) or spherical ($k=1$) universe, this quantity is positive defined. In the hyperbolic space ($k=-1$), always exist a region where the fluid possesses a negative energy density (in particlular, $\rho=-3/(a_0\kappa)^2$ at $t=t_0$), and, as in the previous Section, we will concentrate on the first two cases only. 

The time derivative of the fluid energy density reads\\
\phantom{line}
\begin{equation}
\dot{\rho}=
-\frac{4n(t-t_0)^{2n-3}\alpha[2n(t-t_0)^{2n}\alpha(a_0(1-2n)+(t-t_0)^{2n}\alpha)+k(t-t_0)^2]}
{3(a_0+\alpha(t-t_0)^{2n})^3}\kappa^2\,.
\end{equation}
\phantom{line}\\
When $t$ is close to $t_0$, this expression leads to
\begin{equation}
\dot{\rho}(t\rightarrow t_0)\simeq
\frac{8n^2(t-t_0)^{4n-3}\alpha^2(2n-1)}
{3a_0^2}\kappa^2\,,
\end{equation}
such that the energy density decreases before the bounce and increases after it. However, when $|t|\gg t_0$, one has
\begin{equation}
\dot{\rho}(|t|\gg t_0)=
-\frac{4n(t-t_0)^{-4n-3}[2n(t-t_0)^{4n}\alpha^2+k(t-t_0)^2]}
{3\alpha^2}\kappa^2\,,
\end{equation}
and the energy density increases in the region before (but far from) the bounce and decreases after it. This behaviour could be interesting in the attempt to reproduce the inflation with an alternative scenario with respect to the standard Big Bang one. In the cases of spherical or flat spatial topology, the energy density of the universe starts to decreases after some times from the bounce making possible an exit from inflation. Note that the cosmological paramter reads
\begin{equation}
\Omega=1+\frac{k}{4n^2\alpha^2(t-t_0)^{4n-2}}\,,
\end{equation}
and 
decreases with the acceleration. 

\subsection{Fluids realizing the bounce with power-law scale factor}

In order to reproduce the bounce cosmology with power-law scale factor we need a viscosity in the EoS of the fluids. 
For a generic spatial topology we obtain
\begin{equation}
p=-\frac{\rho}{3}-3H\zeta(a(t), H)\,,
\end{equation}
where
\begin{equation}
\zeta(a(t), H)=\frac{(2n-1)a(t)}{3n(a(t)-a_0)\kappa^2}\,.
\end{equation}
When $a(t)\gg a_0$ the bulk viscosity reads
\begin{equation}
\zeta(H,a(t)\gg a_0)\simeq\frac{(2n-1)}{3n\kappa^2}\,,
\end{equation}
and it is quite a constant. We see that this expression is positive for $n>1/2$ but also for $n<0$, when, as we have seen at the beginning of the Chapter, the fluid realizes the Big Rip scenario. If the viscosity $\zeta$ is such that $0<\zeta<2/3$ (it means, $n>1/2$) the bounce solution can be realized, but if $2/3<\zeta$ the fluid may bring the universe evolution to the Big Rip singularity.

\section{Bounce solutions in modified theories of gravity}

In principle, one may 
encode any modification of gravity in the fluid-like form. In this Chapter, following the first proposal of Ref.~\cite{Odbounce}, we will investigate the case of $f(R)$-gravity realizing the bounce cosmology and whose action (in vacuum) is given by
\begin{equation}
I=\int_{\mathcal{M}} d^4x\sqrt{-g}\left[
\frac{R+f(R)}{2\kappa^2}\right]\,,\label{action}
\end{equation}
where $g$ is the determinant of the metric tensor, $g_{\mu\nu}$, $\mathcal{M}$ is the space-time manifold and 
$f(R)$ is a function of the Ricci scalar $R$ and represents the correction to the Einstein's gravity.
In this Chapter, for the sake of simplicity, we will consider the flat FRW metric only, namely (\ref{metric}) with $k=0$.
In this case, the equations of motion read
\begin{equation}
\rho_{\mathrm{eff}}=\frac{3}{\kappa^{2}}H^{2}\,,
\quad
p_{\mathrm{eff}}=-\frac{1}{\kappa^{2}} \left( 2\dot H+3H^{2} \right)\,,\label{F2}
\end{equation}
where $\rho_{\mathrm{eff}}$ and $p_{\mathrm{eff}}$ are
the effective energy density and pressure of the modified gravity model, namely
\begin{eqnarray}
\rho_{\mathrm{eff}} &\equiv&
\frac{1}{2\kappa^{2}}
\left[ \left( f'(R) R-f(R) \right)-6H^2 f'(R)
-6H\dot{f}'(R)
\right]\,,
\label{rhoeffRG} \\ \nonumber\\
p_{\mathrm{eff}} &\equiv&
\frac{1}{2\kappa^{2}} \Bigl[
-\left( f'(R)R-f(R)\right)+(4\dot{H}+6H^2)f'(R)
+4H \dot f'(R)+2\ddot{f}'(R)
\Bigr]\,.\, \nonumber\\
\end{eqnarray}
Here, the prime denotes the derivative with respect to $R$ and the dot (as usually) is the derivative with respect to the time. Thus, we recover the Friedmann equations (\ref{EOMs}), where the modification of gravity is treated like a fluid whose EoS can be written in the form of~(\ref{start}).
We have many possibilities.
For example, we may take $\omega(\rho_\mathrm{F})=\omega$, where $\omega$ is a constant (usually one chooses $\omega=-1$), and identify
the bulk viscosity as
\begin{eqnarray}
\hspace{-5mm}
B(H, \dot{H}...) &=&
-\frac{1}{2\kappa^{2}} \biggl\{ (1+\omega)(f(R)-R f'(R))
+f'(R)
\left[6H^2(1+\omega)+4\dot{H}\right]\label{ex}
\nonumber \\
\hspace{-5mm}
& &
+ H{\dot{f}}'(R)(4+6\omega)
+2{\ddot{f}}'(R)\biggr\}\,.
\end{eqnarray} 
In order to reconstruct modified gravity models realizing the bounce solutions, one can take the time derivative of $\rho_\text{eff}$,
\begin{equation}
\dot{\rho}_\text{eff}=\frac{1}{2\kappa^2}\left[6H^2\dot f'(R)-12 H\dot H f'(R)-6H\ddot f'(R)\right]\,,
\end{equation}
and insert this expression in the derivative of the first Friedmann-like equation, such that, given the bounce solution, we obtain an equation for $f'(R)$ only,
\begin{equation}
\frac{6H\dot H}{\kappa^2}=
\frac{1}{2\kappa^2}\left[6H^2\dot f'(R)-12 H\dot H f'(R)-6H\ddot f'(R)\right]\,.\label{principe}
\end{equation}
From this equation it is possible to derive the on-shell form of $f'(R)$ and then, by replacing the time with the corresponding expression of the Ricci scalar, get the model $f(R)$ from $f'(R)$.\\
\\
Let us see some examples of modified gravity model realizing the bounce. The Ricci scalar reads
\begin{equation}
R=12H^2+6\dot H\,.
\end{equation}
We can start from solution (\ref{exp}) with $n=1$. The Hubble parameter and the Ricci scalar are given by
\begin{equation}
H(t)=2\alpha(t-t_0)\,,\quad R=48\alpha^2(t-t_0)^2+12\alpha\,.
\end{equation}
The solution of (\ref{principe}) is 
\begin{equation}
f'(R)=c_0\left[1-2\alpha(t-t_0)^2-\frac{1}{c_0}\right]\,,
\end{equation}
namely
\begin{equation}
f'(R)=c_0\left[\left(\frac{3}{2}-\frac{1}{c_0}\right)-\frac{R}{24\alpha}\right]\,,
\quad
f(R)=c_0\left[\left(\frac{3}{2}-\frac{1}{c_0}\right)R-\frac{R^2}{48\alpha}\right]+c_1\,,
\end{equation}
where $c_0, c_1$ are generic constants. By using the first Friedmann-like equation (\ref{F2}), one obtains
\begin{equation}
c_1=-3\alpha c_0\,.
\end{equation}
This result is in agreement with Ref.~\cite{Odbounce}. In order to recover the Einstein gravity term in the action, we must put
\begin{equation}
c_0=\frac{2}{3}\,,\quad
f(R)=-\frac{R^2}{72\alpha}-2\alpha\,.
\end{equation}
Some comments are in order. It is well know that the $\gamma R^2$-term with positive $\gamma>0$~\cite{staro} or $f(R)=\gamma R^2+\lambda$, where $\gamma, \lambda>0$~\cite{mio}
admit accelerated non-singular solutions and are used in the theories for inflation. Here, we see that if the coefficient in front of $R^2$ and the ``cosmological'' constant of the model are negative, we still obtain an accelerated solution, but back into the past the bounce appears.\\

An other example simple to solve is given by (\ref{pow}) with $a_0=0$. In such a case, 
\begin{equation}
H(t)=\frac{2n}{(t-t_0)}\,,\quad R(t)=\frac{48n^2-12n}{(t-t_0)^2}\,,
\end{equation}
and the reconstruction leads to
\begin{equation}
f'(R)=-1+c_0\left(\frac{R}{48n^2-12n}\right)^{\lambda_\pm}\,,\quad f(R)=
-R+\frac{c_0}{(\lambda_\pm+1)}\left(\frac{R}{48n^2-12n}\right)^{\lambda_\pm}R+c_1\,,
\end{equation}
where
\begin{equation}
\lambda_\pm=-\frac{1}{4}\left(1+2n\pm\sqrt{1+20n+4n^2}\right)\,.
\end{equation}
Here, $c_0$ is a free paramter and $c_1$ is fixed by the first Friedmann-like equation as
\begin{equation}
c_1=0\,,
\end{equation}
such that the final Lagrangian of the theory results to be a power-law of the Ricci scalar,
\begin{equation}
\mathcal L=c_0 R^{\lambda_\pm+1}\,,
\end{equation}
where we have redefined the constant $c_0$. If $n<0$, we find a model for the Big Rip~\cite{OdOd}, such that for this kind of modified power law-models the appearance of the Big Rip or the bounce solution must be carefully investigated. Note that in our case, making the choice $a_0=0$ in (\ref{pow}), the Hubble paramter diverges at $t=t_0$ like for the Big Rip: however, the scale factor does not diverge and realizes the bounce.
We also observe that $\lambda_\pm+1\neq 1$ independently on $n$, according with the fact that the pure Einstein gravity is free of bounce or singularity solutions.

\section{Conclusions}

The study of inhomogeneous viscous fluids in Friedmann-Robertson-Walker universe is important 
under many points of view. This kind of fluids has a very general form of Equation of State
and can be used in many different contexts, like the description of current dark energy epoch or the primordial inflation.
Then, many dark energy models have a corresponding fluid-representation, and the modified theories of $f(R)$-gravity are an example of it.

In this paper, we have analyzed the bounce cosmology realized in Friedmann-Robertson-Walker space-time
by viscous fluids considering two specific forms of the bounce, namely bounce with exponential scale factor, and bounce with power-law scale factor. In the both cases, we have found some examples of fluids which bring to such solutions. 
Since the bounce has been proposed as an alternative scenario for the Big Bang, our investigation has taken into account the necessity to have an acceleration after the bounce in the context of inflation. For this reason, we have considered different topologies (non necessarly flat) for the FRW metric and we have payed attention to the evolution of the cosmological parameter $\Omega$ also. Generally speaking, the bounce solutions bring to an accelerated universe, but the behaviour of the related fluid energy densities can be different. It is reasonable to expect a decreasing of the energy density during the contraction phase, and an increasing of it in the expanding universe: however, we have seen that in some case it may exist a region around the bounce where this behaviour is inverted. This fact is quite interesting, since it gives the possibility to reproduce after the bounce an accelerating universe whose energy density decreases, making possible an exit from this stage, but it is clear that to obtain a realistic inflation other universe contents must be considered in the theory in order to bring the universe to a decelerated expansion. 
An other interesting point analyzed in this work is the relation between bounce and singualar solutions: since the form of the scale factors is the same, also the related fluids present the same structure of Equation of State, and the occurrence of one solution instead to the other one typically depends on the coefficients of the bulk viscosity only.

In the final part of the paper, following the first proposal of Ref.~\cite{Odbounce}, we have analyzed $f(R)$-modified gravity realizing the bounce and some explicit examples have been derived and discussed for flat FRW space-time. 

For more detailed analysis on the bounce scenario (in the specific, instabilities and ghosts) see Refs.~\cite{det}.
Other relevant works on inhomogeneous viscous fluids and the dark energy issue have been presented in Refs.~\cite{uno}--\cite{otto}, in Ref.~\cite{B1} for the inflationary scenario and in Ref.~\cite{LittleRip} for viscous fluids applied to the study of Little Rip cosmology.


\end{document}